\documentclass[twocolumn,showpacs,preprintnumbers,amsmath,amssymb]{revtex4}

\usepackage{graphicx}
\usepackage{dcolumn}
\usepackage{bm}

\begin{document}

\title{Random Apollonian Networks}
\author{Tao Zhou$^{1,2}$}
\author{Gang Yan$^{2}$}
\author{Pei-Ling Zhou$^{2}$}
\author{Zhong-Qian Fu$^{2}$}
\author{Bing-Hong Wang$^{1}$}
\email{bhwang@ustc.edu.cn,Fax: +86-551-3603574.}
\affiliation{%
$^{1}$Nonlinear Science Center and Department of Modern Physics,
University of Science and Technology of China, Hefei Anhui,
230026, PR China \\
$^{2}$Department of Electronic Science and
Technology, University of Science and Technology of China,  Hefei
Anhui, 230026, PR China
}%

\date{\today}

\begin{abstract}
In this letter, we propose a simple rule that generates scale-free
networks with very large clustering coefficient and very small
average distance. These networks are called {\bf Random Apollonian
Networks}(RANs) as they can be considered as a variation of
Apollonian networks. We obtain the analytic result of power-law
exponent $\gamma =3$ and clustering coefficient
$C=\frac{46}{3}-36\texttt{ln}\frac{3}{2}\approx 0.74$, which agree
very well with the simulation results. We prove that the
increasing tendency of average distance of RAN is a little slower
than the logarithm of the number of nodes in RAN. Since many
real-life networks are both scale-free and small-world, RANs may
perform well in mimicking the reality. The epidemic spreading
process is also studied, we find that the diseases spread slower
in RANs than BA networks in the early stage of SI process,
indicating that the large clustering coefficient may slower the
spreading velocity especially in the outbreaks.
\end{abstract}

\pacs{89.75.Hc,64.60.Ak,84.35.+i,05.40.-a,05.50+q,87.18.Sn}

\maketitle

Recently, empirical studies indicate that the networks in various
fields have some common characteristics, which inspires scientists
to construct a general model\cite{Reviews}. One of the most
well-known models is Watts and Strogatz's small-world network (WS
network), which can be constructed by starting with a regular
network and randomly moving one endpoint of each edge with
probability $p$ \cite{WS}. Another significant one is Barab\'{a}si
and Albert's scale-free network model (BA network)\cite{BA}. The
BA model suggests that two main ingredients of self-organization
of a network in a scale-free structure are growth and preferential
attachment.

\begin{figure}
\scalebox{0.45}[0.45]{\includegraphics{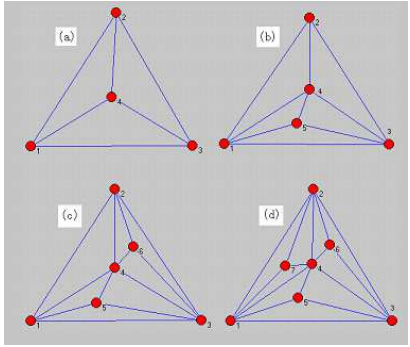}}
\caption{\label{fig:epsart} The sketch maps for the network
growing process. The four figures show a possible growing process
for RAN at $t=1$(a), $t=2$(b), $t=3$(c) and $t=4$(d). At time step
1, the 4th node is added to the network and linked to node 1, 2
and 3. Then, at time step 2, the triangle $\triangle 134$ is
selected, the 5th node is added inside this triangle and linked to
node 1, 3 and 4. After that, the triangles $\triangle 234$ and
$\triangle 124$ are selected  and node 6 and 7 are added inside
them, respectively at time step 3 and 4. Keep on the similar
iterations, one can get RANs of any orders as he like.}
\end{figure}

A few authors have demonstrated the use of pure mathematical
objects and methods to construct scale-free networks. One
interesting instance is the so-called integer networks\cite{IN},
of which the nodes represent integers. Another related work is
owed to Dorogovtsev and Mendes et al, in which the deterministic
networks, named {\bf pseudofractals}, are obtained by attachment
aiming at edges\cite{Dorogovtsev}. Here, we focus on the so-called
{\bf Apollonian Networks}(ANs) introduced by Andrade et
al\cite{AN1}. The networks can be produced as follow: start with a
triangle and then at each generation, inside each triangle, a node
is added and linked to the three vertices. Doye et. al. have
studied the properties of ANs detailedly\cite{AN2}, and shown the
degree distribution $p(k)\propto k^{{\rm -\gamma}}$, average
length $l\propto (ln N)^{{\beta}}$, where $\gamma = 1 + \frac{ln
3}{ln 2} \approx 2.585$, $\beta \approx 0.75$ and $N$ is the
order\cite{ex1}. In this letter, we propose a simple rule that
generates scale-free networks with very large clustering
coefficient and very small average distance. These networks are
called {\bf Random Apollonian Networks}(RANs), since they can be
considered as a variation of ANs.

\begin{figure}
\scalebox{0.7}[0.6]{\includegraphics{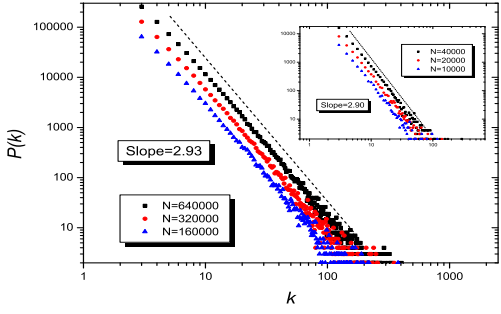}}
\caption{\label{fig:epsart} Degree distribution of RANs, with
$N=640000$(black squares), $N=320000$(red circles) and
$N=160000$(blue triangles), where $P(k)$ denotes the number of
nodes of degree $k$. The power-law exponents are
$\gamma_{640000}=2.94\pm 0.04$, $\gamma_{320000}=2.92\pm 0.05$ and
$\gamma_{160000}=2.92\pm 0.06$, respectively. The average of them
are 2.93. The inset shows Degree distribution with $N=80000$(black
squares), $N=40000$(red circles) and $N=20000$(blue triangles).
The exponents are $\gamma_{640000}=2.91\pm 0.07$,
$\gamma_{320000}=2.90\pm 0.07$ and $\gamma_{160000}=2.90\pm 0.09$.
The mean value is 2.90. The two dash lines have slope -3.0 for
comparison.}
\end{figure}
\begin{figure}
\scalebox{0.5}[0.4]{\includegraphics{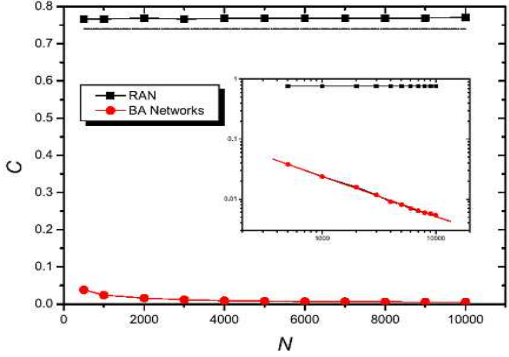}}
\caption{\label{fig:epsart} The clustering coefficient of
RAN(black squares) and BA networks(red circles). In this figure,
one can find that the clustering coefficient of RAN is almost a
constant a little smaller than 0.77, which accords with the
analytic result(dash line) very well. The inset shows that the
clustering coefficient of BA networks, which decreases with the
increasing of network order quickly.}
\end{figure}

RAN starts with a triangle containing three nodes marked as 1, 2
and 3. Then, at each time step, a triangle is randomly selected,
and a new node is added inside it and linked to its three
vertices. The sketch maps for the network growing process are
shown in figure 1. Note that, after a new node is added, the
number of triangles increases by 2. Therefore, we can immediately
get that when the networks are of order $N$, the number of
triangles are:
\begin{equation}
N_{\triangle}=2(N-3)+1=2N-5
\end{equation}
Let $N_{\triangle}^i$ denote the number of triangles containing
the $i$th node, the probability that a newly added node will
connect to the $i$th node is $N_{\triangle}^i/N_{\triangle}$.
Apparently, except the node 1, 2 and 3, $N_{\triangle}^i$ is equal
to the degree of the $i$th node: $N_{\triangle}^i=k_{i}$.
Therefore, we can write down a rate equation for the degree
distribution. Let $n(N,k)$ be the number of nodes of degree $k$
when $N$ nodes are present, we have:
\begin{equation}
n(N+1,k+1)=n(N,k)\frac{k}{N_{\triangle}}+n(N,k+1)(1-\frac{k+1}{N_{\triangle}})
\end{equation}
When $N$ is sufficient large, $n(N,k)$ can be approximated as
$Np(k)$, where $p(k)$ is the probability density function for the
degree distribution. In terms of $p(k)$, the above equation can be
rewritten as:
\begin{equation}
(N+1)p(k+1)=\frac{Nkp(k)}{N_{\triangle}}+Np(k+1)-\frac{N(k+1)p(k+1)}{N_{\triangle}}
\end{equation}
Using Equ.(1) and the expression $p(k+1)-p(k)=\frac{dp}{dk}$, we
can get the continuous form of Equ.(3):
\begin{equation}
k\frac{dp}{dk}+\frac{3N-5}{N}p(k)=0
\end{equation}
This lead to $p(k)\propto k^{-\gamma}$ with
$\gamma=(3N-5)/N\approx 3$ for large $N$. Figure 2 shows the
simulation results, which agree with the analytic one very well .

In succession, let us calculate the clustering coefficient of
RANs. For an arbitrary node $x$, the clustering coefficient $C(x)$
is:
\begin{equation}
C(x)=\frac{2E(x)}{k(x)(k(x)-1)}
\end{equation}
where $E(x)$ is the number of edges among node $x$'s neighbor-set
$A(x)$, and $k(x)=|A(x)|$ is the degree of node $x$. The
clustering coefficient $C$ of the whole network is defined as the
average of $C(x)$ over all nodes. At the very time when node $x$
is added to the network, it is of degree 3 and $E(x)=3$. After, if
the degree of node $x$ increases by one(i.e. a new node is added
to be a neighbor of $x$), then $E(x)$ will increases by two since
the newly added node will link to two of the neighbors of node
$x$. Therefore, we can write down the expression of $E(x)$ in
terms of $k(x)$: $E(x)=3+2(k(x)-3)=2k(x)-3$, which leads to:
\begin{equation}
C(x)=\frac{2(2k(x)-3)}{k(x)(k(x)-1)}
\end{equation}
Consequently, we have:
\begin{equation}
C=\frac{2}{N}\sum_{i=1}^{N}\frac{2k_i-3}{k_i(k_i-1)}=\frac{2}{N}\sum_{i=1}^{N}(\frac{3}{k_i}-\frac{1}{k_i-1})
\end{equation}
where $k_i$ denotes the degree of the $i$th node. Rewrite
$\sum_{i=1}^{N}f(k_i)$ in continuous form, we have:
\begin{equation}
C=6\int_{k_{min}}^{k_{max}}\frac{p(k)}{k}dk-2\int_{k_{min}}^{k_{max}}\frac{p(k)}{k-1}dk
\end{equation}
where $k_{min}$ and $k_{max}$ denote the minimal and maximal
degree in RAN, respectively. Note that $p(k)=\alpha k^{-\gamma}$
with $\gamma =3$ and $\alpha$ a constant, one have:
\begin{equation}
C=6\alpha \int_{k_{min}}^{k_{max}}k^{-4}dk-2\alpha
\int_{k_{min}}^{k_{max}} \frac{1}{k^3(k-1)}dk
\end{equation}
Using the normalization equation
$\int_{k_{min}}^{k_{max}}p(k)dk=1$ and the approximate condition
that $k_{max}\gg k_{min}=3$, we have
$C=\frac{46}{3}-36\texttt{ln}\frac{3}{2}\approx 0.74$.

Figure 3 shows the simulation results about the clustering
coefficient of RAN, which agree very well with the analytic one.
It is remarkable that, the clustering coefficient of BA networks
is very small and decreases with the increasing of network order,
following approximately $C\sim \texttt{ln}^2N/N$\cite{Klemm}.
Since the data-flow patterns show a large amount of clustering in
interconnection networks, the RAN may perform better than BA
networks. In addition, the demonstration exhibits that most
real-life networks have large clustering coefficient no matter how
many nodes they have, which agrees with the case of RAN but
conflicts with the case of BA networks, thus RAN may be more
appropriate to mimic the reality.

At last, let's discuss the average distance of RANs. Marked each
node according to the time when it is added to the network(see
figure 1), then we have the following \textbf{Lemma:} for any two
nodes $i$ and $j$, each shortest path from $i$ to $j$ does not
pass through any nodes $k$ satisfying that
$k>\texttt{max}\{i,j\}$. The proof is a routine exercise thus
omitted here.

Using symbol $d(i,j)$ to represent the distance between $i$ and
$j$, the average distance of RAN with order $N$, denoted by
$L(N)$, is defined as: $L(N)=\frac{2\sigma (N)}{N(N-1)}$, where
the total distance is: $\sigma (N)=\sum_{1\leq i<j\leq N}d(i,j)$.
\begin{figure}
\scalebox{0.7}[0.6]{\includegraphics{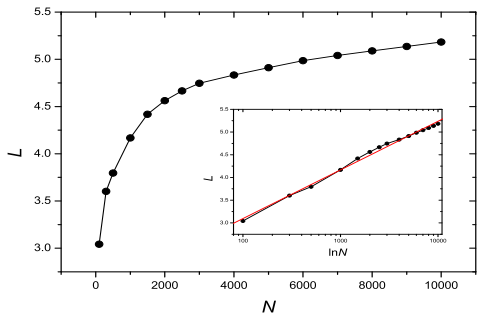}}
\caption{\label{fig:epsart} The dependence between the average
distance $L$ and the order $N$ of RANs. The inset exhibits the
curve where $L$ is considered as a function of $\texttt{ln}N$,
which is well fitted by a straight line. The curve is above the
fitting line when $N$ is small($1000\leq N\leq 3000$) and under
the fitting line when $N$ is large($N\geq4000$), which indicates
that the increasing tendency of $L$ can be approximated as
$\texttt{ln}N$ and in fact a little slower than $\texttt{ln}N$.
All the data are obtained by 10 independent simulations.}
\end{figure}

According to the lemma, newly added node will not affect the
distance between old nodes. Hence we have:
\begin{equation}
\sigma(N+1)=\sigma(N)+\sum_{i=1}^N d(i,N+1)
\end{equation}
Assume that the node $N+1$ is added into the triangle $\triangle
y_1y_2y_3$, then the Equ.(10) can be rewritten as:
\begin{equation}
\sigma(N+1)=\sigma(N)+\sum_{i=1}^N (D(i,y)+1)
\end{equation}
where $D(i,y)=\texttt{min}\{d(i,y_1),d(i,y_2),d(i,y_3)\}$.
Constrict $\triangle y_1y_2y_3$ continuously into a single node
$y$, then we have $D(i,y)=d(i,y)$. Since
$d(y_1,y)=d(y_2,y)=d(y_3,y)=0$, the Equ.(11) can be rewritten as:
\begin{equation}
\sigma(N+1)=\sigma(N)+N+\sum_{i\in \Gamma} d(i,y)
\end{equation}
where $\Gamma =\{1,2,\cdots ,N\}-\{y_1,y_2,y_3\}$ is a node set
with cardinality $N-3$. The sum $\sum_{i\in \Gamma} d(i,y)$ can be
considered as the total distance from one node $y$ to all the
other nodes in RAN with order $N-2$. In a rough version, the sum
$\sum_{i\in \Gamma} d(i,y)$ is approximated in terms of $L(N-2)$:
\begin{equation}
\sum_{i\in \Gamma} d(i,y)\approx (N-3)L(N-2)
\end{equation}
Note that, the average distance $L(N)$ increases monotonously with
$N$, it is clear that:
\begin{equation}
(N-3)L(N-2)=\frac{2\sigma(N-2)}{n-2}<\frac{2\sigma(N)}{N}
\end{equation}
Combining Equs.(12), (13) and (14), one can obtain the inequality:
\begin{equation}
\sigma(N+1)<\sigma(N)+N+\frac{2\sigma(N)}{N}
\end{equation}
Consider (15) as an equation, then the increasing tendency of
$\sigma(N)$ is determined by the equation:
\begin{equation}
\frac{d\sigma(N)}{dN}=N+\frac{2\sigma(N)}{N}
\end{equation}
This equation leads to
\begin{equation}
\sigma(N)=N^2\texttt{ln}N+H
\end{equation}
where $H$ is a constant. As $\sigma(N)\sim N^2L(N)$, we have
$L(N)\sim \texttt{ln}N$. Which should be pay attention to, since
(15) is an inequality indeed, the precise increasing tendency of
$L$ may be a little slower than $\texttt{ln}N$.

\begin{figure}
\scalebox{0.4}[0.5]{\includegraphics{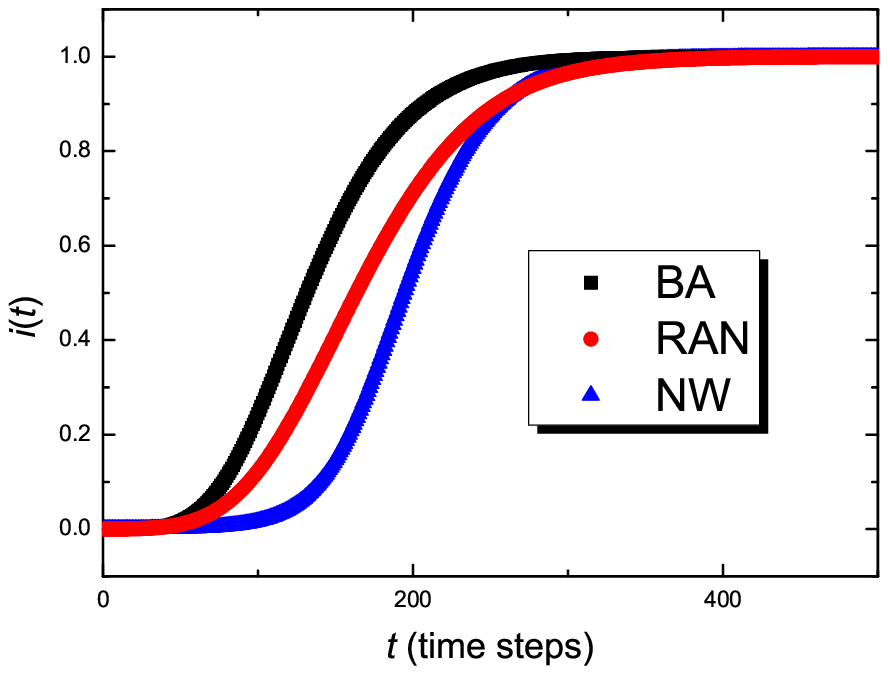}}
\scalebox{0.4}[0.5]{\includegraphics{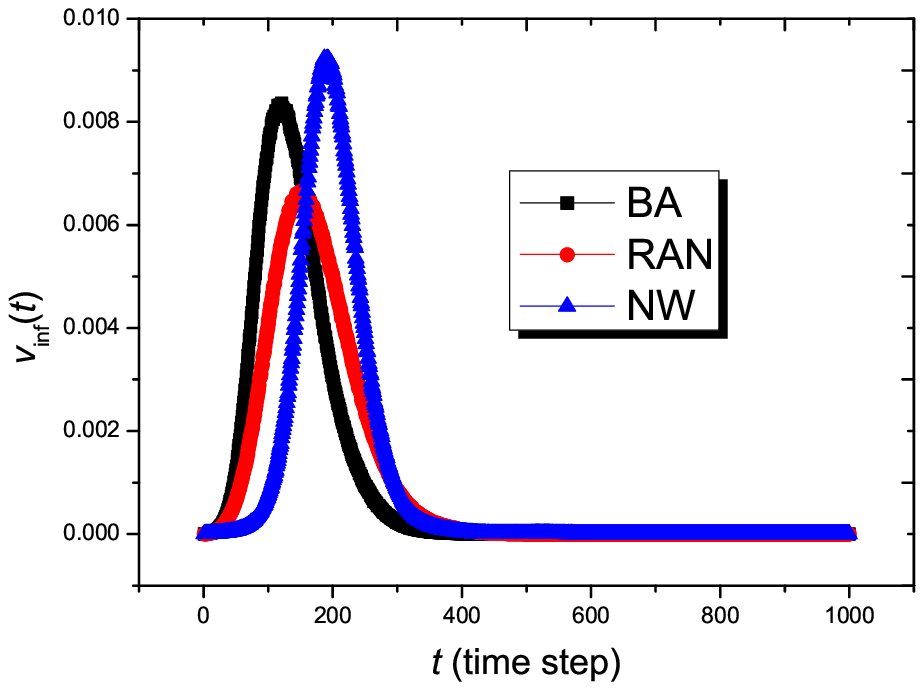}}
\caption{\label{fig:epsart} The left and right plots show average
density of infected individuals and spreading velocity versus time
with $N=10000$ and average degree $\langle k\rangle =6$ fixed. The
black, red and blue curves correspond to the case of BA, RAN and
NW networks respectively. The NW networks are of $z=1$ and $\phi
=4\times 10^{-4}$, thus $\langle k\rangle \approx 2z+\phi
N=6$\cite{ex3}(see also the accurate definitions of $z$ and $\phi$
in reference\cite{Newman1999b}). The spreading rate is
$\lambda=0.01$. All the data are averaged over $10^3$ independent
runs. The spreading velocity reaches a peak quickly. Before the
peak-time, the spreading velocities of the three kinds of networks
satisfy the inequality $v_{inf}^{BA}>v_{inf}^{RAN}>v_{inf}^{NW}$.}
\end{figure}

As we mentioned above, close to many real-life networks, random
Apollonian networks are both scale-free and small-world.
Therefore, it is worthwhile to investigate the processes taking
place upon RAN and directly compare these results with just
small-world and just scale-free networks. One widely studied
process is the epidemic spreading process\cite{Pastor-Satorras};
and for the sake of protecting networks and finding optimal
strategies for the deployment of immunization resources, it is of
practical importance to study the dynamical evolution of the
outbreaks\cite{Barthelemy04,Yan05}. Numerical simulations aiming
at BA, NW\cite{Newman1999a} and RAN are shown in figure 5 with the
process and the conception of density and velocity all the same as
previous studies\cite{Barthelemy04,Yan05}. The result that
diseases spread more quickly in RAN and BA networks than in NW
networks is easy to be understood as the well-known conclusion:
boarder degree distribution will speed up the epidemic spreading
process\cite{Pastor-Satorras}.

Why the diseases spread more quickly in BA networks than RAN is a
very interesting question. We argue that the larger clustering
coefficient may slow down the spreading process especially in the
outbreaks. Remove an arbitrary edge $e(x,y)$ from quondam network,
then, the distance between $x$ and $y$ will increase as
$d'(x,y)>1$(if the removal of $e$ makes $x$ and $y$ disconnected,
then we set $d'(x,y)=N$). The quantity $d'(x,y)$ can be considered
as edge $e$'s score $s(e)=d'(x,y)\geq 2$, denoting the number of
edges the diseases must pass through from $x$ to $y$ or form $y$
to $x$ if they do not pass across $e$. If $s(e)$ is small, then
$e$ only plays a local role in the spreading process, else when
$s(e)$ is large, $e$ is of global importance. For each edge $e$,
if it does some contribution to clustering coefficient, it must be
contained in at least one triangle and $s(e)=2$. Therefore,
networks of larger clustering coefficient have more {\bf local
edges}. RANs and BA networks are two extreme cases of scale-free
networks. In RANs, all the edges are of score 2; while in BA
networks, almost all the edges are of score larger than 2 because
the clustering coefficient of BA networks will decay to zero
quickly as $N$ increases. Consequently, diseases spread more
quickly in BA networks than RAN. The above explanation is
qualitative and rough, to study the process upon networks with
tunable clustering coefficient\cite{Holme} may be useful, which
will be the future work.

In respect that the RANs are of very large clustering coefficient
and very small average distance, they are not only scale-free, but
also small-world. Since many real-life networks are both
scale-free and small-world, RAN may perform better in mimicking
reality than WS and BA networks. The epidemic spreading process is
also studied, in which the diseases spread slower in RAN than BA
networks during the outbreaks, indicating that the large
clustering coefficient may slower the spreading velocity
especially in the outbreaks. This numerical study suggests that
the clustering structure may affect the dynamical behavior upon
networks much. Further more, many real-life networks are planar
networks by reason of technical or natural requirements, such as
layout of printed circuits, river networks upon the earth's
surface, vas networks clinging to cutis, and so forth. Since the
number of edges in RAN is equal to $3N-6$, RAN are maximal planar
networks\cite{Bollobas1998}, which are possibly of particular
practicability for layout of printed circuits and so on.

This work has been partially supported by the National Natural
Science Foundation of China(No. 70171053, 70271070, 70471033 and
10472116), the Specialized Research Fund for the Doctoral Program
of Higher Education (SRFDP No.20020358009) and the Foundation for
Graduate Students of University of Science and Technology of China
under Grant No. KD200408.


\begin{thebibliography}{Reviews}
\bibitem{ex1} In this paper, the terminology ``order" means the number of nodes in
network, which is also called network size in some other papers.
\end{thebibliography}
\end{document}